\documentstyle[12pt]{article}
\input{epsf}
\begin{document}
\title {Nonet Classification of Scalar/Isoscalar Resonances
below 1900 MeV: the Existence of an Extra Scalar State in the Region
1200-1600 MeV}
\author{V.V. Anisovich$^{a1}$, Yu.D. Prokoshkin$^{b2}$,
and A.V. Sarantsev$^{a3}$}
\date{}
\maketitle
\newcommand{\be}{\begin{eqnarray}}
\newcommand{\ee}{\end{eqnarray}}
\newcommand{\inli}{\int\limits}

\begin{abstract}

A classification of the  $IJ^{PC}=00^{++}$  mesons
is performed on the basis of the K-matrix analysis of
meson spectra in the reactions:
(i) GAMS data on $\pi p\to \pi^0\pi^0 n$,
$\eta\eta n$, $\eta\eta' n$ \cite{gams1,gams2,gams3}; (ii) Crystal
Barrel data on $p\bar p~(at~rest)\to \pi^0\pi^0\pi^0$, $\pi^0\pi^0\eta$,
$\pi^0\eta\eta$ \cite{cbc1,cbc2};
(iii) CERN-M\"unich data on $\pi p\to \pi^+\pi^- n$ \cite{cern};
(iiii) BNL data on
$\pi N\to K\bar K N$ \cite{bnl}.
The analysis points to
the existence of four comparatively narrow scalar resonances
which correspond to the following poles of the scattering
amplitude (in MeV):
$(1015\pm 15)- i(43\pm 8)$, $(1300\pm 20)-i(120\pm 20)$,
$(1499\pm 8)-i(65\pm 10)$ and $(1780\pm 30)-i(125\pm 70)$.
The scattering amplitude also has a fifth pole
$f_0(1530^{+90}_{-250})$ at the
complex mass $(1530^{+90}_{-250}) - i(560\pm 140)$. The
masses of the K-matrix poles (bare states) are at $720\pm 100$ MeV,
$1230\pm 50$ MeV, $1260\pm 30$ MeV, $1600\pm 50$ MeV and
$1810\pm 30$ MeV. The quark content of the bare states is
analyzed using the values of their couplings to the
$\pi\pi$, $K\bar K$, $\eta\eta$ and $\eta\eta'$.
It is shown that one of the bare states in the mass region 1200-1600 MeV
is superfluous for the $q\bar q$
classification and should be considered as the lightest glueball.
\end{abstract}

\vspace{4cm}

{\parindent=0cm
\line(1,0){290}
\footnotesize

a) {\it St.Petersburg Nuclear Physics Institute, Gatchina, 188350 Russia}\\
b) {\it Institute for High Energy Physics, Protvino, 142284 Russia}  \\
1) anisovic@lnpi.spb.su,  \ anisovic@thd.pnpi.spb.ru\\
2) prokoshkin@mx.ihep.su, \ Yuri.Prokoshkin@cern.ch\\
4) vsv@hep486.pnpi.spb.ru, \  andsar@v2.rl.ac.uk }

\newpage

The search for and classification of scalar/isoscalar
$IJ^{PC}=00^{++}$ states is the direct and possibly the only way
for identification of the lightest scalar glueball.
In refs. \cite{aps,kmatr} the K-matrix analysis of the
$00^{++}$-wave was performed in the mass region up to 1550 MeV, where
four scalar resonances (the T-matrix poles at the complex masses:
$(1008\pm 10)-i(43\pm 5)$, $(1290\pm 25)-i(120\pm 15)$,
$(1497\pm 6)-i(61\pm 5)$, $(1430\pm 150)-i(600\pm 100)$, in MeV)
were found.
Correspondingly, four bare states were determined: the
lightest bare state with mass $750\pm 100$ MeV is
dominantly $s\bar s$, while three other states, with masses
$1240\pm 30$ MeV, $1280\pm 30$ MeV and
$1615\pm 50$ MeV, do not contain a large $s\bar s$ component.
One of these states, either the state with
mass 1240 MeV or the state with mass 1280 MeV, is a natural
$q\bar q$-partner of the lightest bare state. For the other
two states two scenarios arose in ref. \cite{kmatr}:
\begin{description}
\item[(a)] both these states are $q\bar
q$ mesons; then, in the region 1550-1900 MeV, two $s\bar s$-rich
states exist as the nonet partners of the low-lying
$00^{++}$-mesons;
\item[(b)] in the region 1550-1900 MeV there
is only one $s\bar s$-rich $00^{++}$ state; then one of the
low-lying states is exotic, probably  the lightest glueball.
\end{description}

To resolve these alternatives, the spectra in $K\bar K$, $\eta\eta$ and
$\eta\eta'$ channels need to be investigated in the region
1550-1900 MeV: the existence of a strange component in these mesons
favours a search for the $s\bar s$-rich states.
The $\eta\eta$ and $\eta\eta'$ spectra obtained by the GAMS collaboration
\cite{gams2,gams3} give a good opportunity  for this study. The aim here
is to extend the analysis of the $00^{++}$-wave to a mass of
1900 MeV, including the $\eta\eta$ and $\eta\eta'$ GAMS spectra
into the simultaneous fitting procedure. The main purpose  is to
identify  the $s\bar s$-rich states: the analysis of the $\pi^0\pi^0$ and
$\pi^+\pi^-$ spectra from refs.  \cite{kmatr,bsz} definitely indicates
that in the region 1550-1900 MeV there are no $00^{++}$ resonances
with a significant
$\pi\pi$ branching ratio: so, the presence of $n\bar n$-dominant states is
unlikely here. As to $s\bar s$-rich states, the
radiative $J/\psi$ decays hint at the possible existence of a
scalar resonance near 1750 MeV \cite{pdg}.

Lattice QCD calculations predict the mass of the purely gluonic $0^{++}$
state (glueball) in the region 1500-1750 MeV:
$1550\pm 50$ MeV \cite{ukqcd} and $1707\pm 64$ MeV \cite{ibm}.
However, if the glueball is near 1500  MeV,
it must definitely include  quark degrees of freedom, mainly the
$q\bar q$-component. Quark-antiquark
loop diagrams (Fig. 1a) will reduce the mass of a pure glueball:
the mass shift is of the order of 100-300 MeV \cite{aas}. Another
source of the glueball--mass shift is its possible mixture with
neighbouring $q\bar q$-mesons. The last of these effects can be
taken into account by working within the K-matrix technique.

The advantage of the K-matrix approach is its ability to analyze
the structure of multichannel partial amplitudes
of overlapping resonances.  The K-matrix amplitude is unitary and
correctly represents analytic properties on the right-hand side of
the complex $s$-plane. Although this approach does not reproduce the
analytic structure of the amplitude on the left-hand side of the s-plane,
one may hope that this does not lead to significant inaccuracy in
finding the pole positions and coupling constants in the mass region
under consideration. Left-hand side singularities of the partial amplitude
can be taken into account in the framework of the multichannel
dispersion relation N/D method: we consider this approach
as a necessary though later step in the analysis of the $00^{++}$
amplitude.

 K-matrix analysis demonstrates \cite{aps,kmatr} that poles of
the partial amplitude (or physical poles which correspond to the
observable states) are determined by the mixture of  input
states related to the K-matrix poles via their transition into
real mesons. The wave function of a physical
state is a mixture of not only the input states but also of
real mesons, which realize this mixture and are responsible for the
decay of the physical state. Because of this phenomenon, we call
the input states  "bare states", i.e. the states without a cloud
of real meson. Decay coupling constants of  bare states
are fixed by their quark-gluon content \cite{gersh,amsl,anis1}.
So bare states can be classified by means of their couplings
to different two-meson channels. We perform here such a
classification of the bare states, $f_0^{bare}$, using the
ratios of their coupling constants to the $\pi\pi$, $K\bar K$,
$\eta\eta$, and $\eta\eta'$ channels in the leading terms of the
$1/N_c$ expansion \cite{thooft} (however, for the candidates for a
glueball, the-next-to-leading terms will also be estimated). Our
analysis gives  evidence for the existence of two $q\bar q$-nonets,
while one bare state with mass around 1200-1600 MeV is  superfluous in the
$q\bar q$-systematics. So, the analysis points to the scenario (b),
limiting the mass of the exotic state to the range 1200-1600 MeV.
Large coupling constants indicate that this superfluous bare
state is dispersed
over neighbouring physical states: the narrow $f_0(1300)$ and
$f_0(1500)$ resonances and the broad $f_0(1530^{+90}_{-250})$.

\vskip 0.5cm
{\bf 1) K-matrix approach and quark combinatoric rules
for the decay coupling constants}\\

The standard K-matrix technique is used for the description
of the meson scattering amplitudes in the $00^{++}$-channel:
\be
A= K(I-i\rho K)^{-1},
\ee
where $K_{ab}$ is a 5$\times $5 matrix ($a,b$ = 1,2,3,4,5), with the
following notations for meson states: 1 = $\pi\pi$, 2 = $K\bar K$,
3 = $\eta\eta$ 4 = $\eta\eta'$ and 5 = $\pi\pi\pi\pi$ +other
multimeson states.
The phase space matrix is diagonal: $\rho_{ab}=\delta_{ab}
\rho_a$, with the following $\rho_a$:
$$
\rho_a(s)=\sqrt{\frac{s-4m_a^2}{s}}\qquad , \qquad a=1,2,3,
$$
\be
\rho_4(s)=\left \{
\begin{array}{cl}
\sqrt{\left (1-\frac{(m_\eta+m_{\eta'})^2}{s}\right )
\left (1-\frac{(m_\eta-m_{\eta'})^2}{s}\right )} &
,\;\; s>(m_\eta-m_{\eta'})^2\\
0 &,\;\; s<(m_\eta-m_{\eta'})^2\;\; .
\end{array} \right .
\ee
Here $m_1=m_\pi$, $m_2=m_K$, $m_3=m_\eta$.
Phase space
factors are responsible for the threshold singularities of the
amplitude: to prevent the violation of  analytic properties
we use analytic continuation for $\rho_a$ below threshold.
For example, the $\eta\eta$ phase space factor
$\rho_{a}=(1-4m_\eta^2/s)^{1/2}$ becomes equal
to $i(4m_\eta^2/s-1)^{1/2}$  below the $\eta\eta$ threshold. The phase space
factors we use lead to false kinematic singularities at
$s=0$ (in all phase space factors) and at $s \leq (m_{\eta'}-m_\eta)^2$
(in the $\eta\eta'$ space factor); but these false singularities (which
are standard for the K-matrix approach) are rather distant from the
investigated physical region. For multimeson phase volume at $s$ below
1 GeV$^2$, we use the four-pion phase space defined either as
$\rho \rho$ phase space or as $\sigma \sigma$ phase space.
The result is
practically the same in the two cases and we show the $\rho\rho$
phase space, for which formulae are simpler:
\be
\rho_5(s)=\left \{ \begin{array}{cl}
\rho_0\int\frac{ds_{1}}{\pi}\int\frac{ds_{2}}{\pi}
\frac{M^2\Gamma(s_{1})\Gamma(s_{2})
\sqrt{(s+s_{1}-s_{2})^2-4ss_{1}}}
{s[(M^2-s_{1})^2+M^2\Gamma^2(s_{1})]
[(M^2-s_{2})^2+M^2\Gamma^2(s_{2})]} &,\;\; s<1\;GeV^2\\
1 &,\;\; s>1\;GeV^2\;\;.  \end{array} \right .
\ee
Here $s_1$ and $s_2$ are the two-pion energies
squared, $M$ is the $\rho$-meson mass and $\Gamma(s)$ is its energy-dependent width,
$\Gamma(s)=\gamma\rho_1^3(s)$. The factor $\rho_0$ provides
continuity of $\rho_5(s)$  at $s=1$ GeV$^2$.

For  $K_{ab}$  we use the parametrization similar to that of
ref. \cite{aps}:
\be
K_{ab}(s)=\left ( \sum_\alpha \frac{g^{(\alpha)}_a
g^{(\alpha)}_b}
{M^2_\alpha-s}+f_{ab}\frac{1+s_0}{s+s_0} \right )\;
\frac{s-m_\pi^2/2}{s}\;\;,
\ee
where $g^{(\alpha)}_a$ is a coupling constant of the bare state
$\alpha$ to the meson channel; the parameters $f_{ab}$ and $s_0$
describe a smooth part of the K-matrix elements ($s_0>1.5$ GeV).

The following formulae describe the GAMS $\pi\pi$, $\eta\eta$ and
$\eta\eta'$ production amplitude due to t-channel exchange:
$$
A_{\pi N\to N b}=
N(\bar\Psi_N\gamma_5\Psi_N)F_N(t)D(t)
\tilde K_{\pi\pi(t),a}(1-i\rho  K)^{-1}_{ab}\;\;,\quad
b=\pi\pi,\eta\eta,\eta\eta'\;\;,
$$
\be
\tilde K_{\pi\pi(t),a}=\left (
\sum_\alpha \frac{\tilde g^{(\alpha)}(t)
g^{(\alpha)}_a} {M^2_\alpha-s}+\tilde f_{a}(t)\;
\frac{1+s_0}{s+s_0} \right )\;\frac{s-m_\pi^2/2}{s}\;\;,
\ee
$$
\tilde g^{(\alpha)}(t)=
g^{(\alpha)}_1+(1-\frac{t}{m_\pi^2})\,
(\Lambda_g-\frac{t}{m_\pi^2})g'^{(\alpha)}\;\;,
$$
\be
\tilde f_{a}(t)=f_{1a}+(1-\frac{t}{m_\pi^2})
(\Lambda_g-\frac{t}{m_\pi^2})f'_{a}\;\;,
\ee
\be
F_N(t)=\left [ \frac{\tilde\Lambda-m_\pi^2}
{\tilde\Lambda-t}\right ]^4\;\;,
\qquad D(t)=(m_\pi^2-t)^{-1}\;\;.
\ee
Here $F_N(t)$ is
the nucleon form factor and $D(t)$ is the pion propagator.

The part of the amplitudes $p\bar p~(at~rest)\to \pi^0\pi^0\pi^0$,
$\pi^0\eta\eta$ which corresponds to the two-meson
production in $00^{++}$ state
$A_{p\bar p\to mesons}=A_1(s_{23})+A_2(s_{13})+A_3(s_{12})$
(where the amplitude $A_k(s_{ij})$ stands for any interaction of
particles in intermediate states but with last interactions
when the particle $k$ is a spectator) has the following form:
$$
A_1(s_{23})=\tilde K_{p\bar p\pi,a}
(s_{23})\left(1-i \rho K\right)^{-1}_{ab}\;\;,
\quad b=\pi\pi,\eta\eta\;\;,
$$
\be
\tilde K_{p\bar p\pi,a}(s_{ij})=\left (
\sum_\alpha \frac{\Lambda^{(\alpha)}_{p\bar p\pi} g^{(\alpha)}_a}
{M^2_\alpha-s_{ij}}+\phi_{p\bar p\pi,a}\;
\frac{1+s_0}{s_{ij}+s_0} \right)\;
\frac{s_{ij}-m_\pi^2/2}{s_{ij}}\;\;.
\ee
The same part of the amplitude in the
$p\bar p~(at~rest)\to \pi^0\pi^0\eta$ reaction is described as:
$$
A_1(s_{23})=\tilde K_{p\bar p\eta,a}
(s_{23})\left(1-i \rho K\right)^{-1}_{ab}\;\;,\quad b=\pi\pi\;\;,
$$
\be
\tilde K_{p\bar p\eta,a}(s_{ij})=\left (
\sum_\alpha \frac{\Lambda^{(\alpha)}_{p\bar p\eta} g^{(\alpha)}_a}
{M^2_\alpha-s_{ij}}+\phi_{p\bar p\eta,a}\;
\frac{1+s_0}{s_{ij}+s_0} \right)\;
\frac{s_{ij}-m_\eta^2/2}{s_{ij}}\;\;.
\ee
Parameters $\Lambda^\alpha_{p\bar p\pi}$ and $\phi_{p\bar p\pi}$
($\Lambda^\alpha_{p\bar p\eta}$, $\phi_{p\bar p\eta}$)
may be complex magnitudes with different phases due to three particle
interactions.

In the leading terms of the $1/N$ expansion, the couplings of the
$q\bar q$-meson and glueball to the two
mesons are determined
by the diagrams where  gluons produce $q\bar q$-pairs (see Figs. 1b, c).
The production of soft $q\bar q$ pairs by
gluons violates flavour symmetry: the direct
indication of such a violation comes from  the description of the
multiparticle production
in the central hadron collisions at high energies
(see ref. \cite{ahkm} and references therein) and from
radiative $J/\psi$-decays \cite{anis1}. In these cases the production of
strange quarks is suppressed by the same factor $\lambda$. The ratios of the
production probabilities are $u\bar u:d\bar d:s\bar s=1:1:\lambda$, with
$\lambda=0.4-0.5$ \cite{ahkm}, that makes it possible to calculate
unambigously
the ratios of the decay coupling constants in the framework of the quark
combinatoric rules.  Previously,  quark combinatorics were successfully
applied to the calculation of the hadron production in high energy collisions
\cite{ansh,bjfa} and in the $J/\psi$-decay \cite{vol}.  Extending this
property to the decays of $00^{++}$ $q\bar q$-mesons, one may calculate the
ratios of  coupling constants $f_0\to\pi\pi$, $K\bar K$, $\eta\eta$,
$\eta\eta'$, $\eta'\eta'$. They are given in Table 1 for $f_0=n\bar
n\;\cos\Phi+s\bar s\;\sin\Phi$, where $n\bar n=(u\bar u+d\bar d)/\sqrt 2$.

The glueball decay couplings in the leading terms of $1/N$-expansion
obey the same ratios as the $q\bar
q$-meson couplings, with the  mixing angle $\Phi=\Phi_{glueball}$,
where $\tan\Phi_{glueball}=\sqrt{\lambda/2}$ \cite{kmatr}.
This is  resulted
from the two-stage decay of a glueball (see Fig. 1c):
intermediate $q\bar q$-state in the glueball decay is a mixture
of  $n\bar n$ and $s\bar s$ quarks, with the angle
$\Phi_{glueball}=25^o\pm 3^o$.

In Table 1 we also present  the glueball couplings in the scheme of
Fig. 1d: these couplings are suppressed by the factor $1/N_c$  as compared
to that of Fig. 1c. Nevertheless, we take them into account in  the
analysis of $f_0^{bare}$ considered here as  candidates for a
glueball. The normalization in Table 1 is done,  following
ref. \cite{aas}, in such a way that
the sum of the couplings squared over all channels is proportional to the
probability of the production of new quark pair, $(2+\lambda)$:
\be
\sum\limits_{channels}
G^2(c)I(c)=\frac{1}{2}G^2(2+\lambda)^2,\qquad\qquad
\sum\limits_{channels}
g_G^2(c)I(c)=\frac{1}{2}g_G^2(2+\lambda)^2.
\ee
Here $I(c)$ is the identity factor and $c=\pi^0\pi^0$, $\pi^+\pi^-$,
$K^+K^-$ and so on (see Table 1). With this
normalization $ g_G/G\simeq1/N_c$.
Our experience of quark-gluon diagram calculations teaches us that
the factor $1/N_c$ actually leads  to a suppression of
the order of $1/10$: in the fitting procedure we impose a
restriction $| g_G/G|<1/3$.

We use the coupling constant ratios shown in Table 1 for the
determination of the quark/gluonic content of $f_0^{bare}$.
Justification of this procedure is seen in the multichannel
$N/D$-method: the couplings of $f_0^{bare}$ satisfy the
same ratios as the decay couplings of resonances in the
dispersion relation approach \cite{aas}.

\vskip 0.5cm
{\bf 2) Fit of the data}

The fitting procedure used here is the same as in ref. \cite{kmatr}.
Complications are due to the additional channel,
$\eta\eta'$, and to the new K-matrix pole near 1800 MeV. We investigate a
necessity for this fifth pole, fitting the data with and without it.
The result
is that for the description of the $\eta\eta$ and $\eta\eta'$
spectra above 1700 MeV, the K-matrix pole at 1800 MeV is
definitely needed.
We check the two-pole structure of the
K-matrix elements in the range 1200-1400 MeV, performing the fits in the
two- and one-pole approximations. The results  confirm the
statement of ref. \cite{kmatr}:  the K-matrix without two-pole
structure fails to describe data in the region 1100-1500 MeV.
The one-pole approximation
does not give a satisfactory description of either the Crystal Barrel or
GAMS data on the $\pi^0\pi^0$ spectra at large momentum
transfer squared, $t$. The latter show a well defined peak at
1300 MeV which corresponds to the $f_0(1300)$ resonance (see
Fig. 6).

In ref. \cite{kmatr} two types  of solution were  found.
In the present analysis, which covers the region of higher masses,
up to 1900 MeV, there also exist two groups of solutions which are
actually the continuations of solutions obtained in \cite{kmatr}.
In solution {\bf I}, the  mixing angle $\Phi(1810)$ is
positive and the resonance $f_0(1780)$ is  narrow:
$\Gamma(1780)=140\pm 20$ MeV.
In solution {\bf II} $\Phi(1820)$ is negative
while $\Gamma(1780)=310\pm 50$ MeV. Let us stress that the $00^{++}$
resonance in the region 1750-1800 MeV was seen in the four-pion
system in the decay $J/\psi\to\gamma 4\pi$, and two
different solutions also give either a narrow \cite{jpsi1}  or a broad
\cite{jpsi2} resonance, just as obtained here.

Our nonet classification will be based on the following two constraints:
\begin{description}
\item[(1)] the angle difference between nonet partners should be
$90^o$; for this value the corridor $90^o\pm 5^o$ is allowed in our analysis.
\item[(2)] coupling constants $g$ of  Table 1 should be
approximately equal to each other for nonet partners.
\end{description}
The conventional quark model requires the equality of the
coupling constants $g$.
But the energy dependence of the loop diagram of Fig. 1a, $B(s)$,  may
violate this coupling constant equality because of mass differences of
the nonet partners.
Coupling constants of the K-matrix contain an additional
$s$-dependent factor as compared to the couplings obtained in the
N/D-method: $g^2(K) \to g^2(N/D)/(1+B'(s))$ \cite{aas}. The factor
$(1+B'(s))^{-1}$ affects mostly the low-$s$ region due to the threshold
and left-hand side singularities of the partial amplitude.
Therefore, the coupling constant equality is mostly violated for the
lightest $00^{++}$ nonet, $1^3P_0$ $q\bar q$. We allow for the members
of this nonet $1\leq g(1)/g(2)\leq 1.5$, where the notations 1 and
2 refer to different $f_0^{bare}$.
For the $2^3P_0$ $q\bar q$
nonet members, we put $g(1)/g(2)=1$.

In solution {\bf{I}} the following variant of the nonet
classification exists:
\begin{description}
\item[I.] $f_0^{bare}(720)$ and
$f_0^{bare}(1260)$ are $1^3P_0$ nonet partners,\\ $f_0^{bare}(1600)$ and
$f_0^{bare}(1810)$ are $2^3P_0$ nonet partners,\\
 $f_0^{bare}(1230)$ is a glueball.
\end{description}

For this variant the
$\chi^2$ values are given in the
second column of Table 2, parameters are presented in
Table 3 and the description of data
is shown by dashed curves in Figs. 2-6.

Within solution {\bf{II}}, two variants
describe well the data set:
\begin{description}
\item[II-1.] $f_0^{bare}(720)$ and $f_0^{bare}(1260)$ are $1^3P_0$
nonet partners,\\ $f_0^{bare}(1600)$ and $f_0^{bare}(1810)$ are
$2^3P_0$ nonet partners,  \\ $f_0^{bare}(1230)$ is a glueball;
\item[II-2.] $f_0^{bare}(720)$ and $f_0^{bare}(1260)$ are $1^3P_0$
nonet partners, \\$f_0^{bare}(1230)$ and $f_0^{bare}(1810)$ are
$2^3P_0$ nonet partners,\\ $f_0^{bare}(1600)$ is a glueball.
\end{description}

The $\chi^2$ values for the solutions {\bf{II-1}} and {\bf{II-2}}
are given in
the third and fourth columns of Table 2. Parameters
are presented in Table 4 (solution {\bf{II-1}}) and in Table 5 (solution
{\bf{II-2}}) and the description of data is shown in Figs. 2-5 by
dotted curves for the solution {\bf{II-1}} and in Figs. 2-8 by solid
curves for the solution {\bf{II-2}}.

The t-dependent couplings obtained from  GAMS data
and the production constants  for the solution {\bf{II-2}}
for the Crystal Barrel data are presented in Table 6.

In all the solutions the calculated branching ratios,
$p\bar p \to 3\pi ^0/\eta \eta \pi$, for the description of
Crystal Barrel data are very close to the
experimental value $3.2\pm0.8$ \cite{cbc2}:

\begin{center}
BR$(p\bar p \to \pi \pi \pi/\eta \eta \pi)=$2.85({\bf{I}}),
2.72({\bf{II-1}}), 2.80({\bf{II-2}}).
\end{center}

The imposing of combinatoric rules
on the resonance coupling constants
and nonet classification constraints does not significantly
change the description of data as compared to the fit with free
couplings. Moreover, the use of quark combinatorics
provides a good convergence to the fit, whereas the fit with free
couplings has rather poor convergence, thus yielding serious problems
in finding the main minimum of $\chi^2$.

\vskip 0.5cm
{\bf 3) Results}

Our simultaneous K-matrix analysis of the $00^{++}$-wave points to the
existence of five  bare states $f_0^{bare}$ in the mass region
below 1900 MeV.
Only two of them are definitely $s\bar s$-rich states:
$f_0^{bare}(720)$ and $f_0^{bare}(1810)$. Therefore only two
nonets below 1900 MeV can be constructed; the following variants of the
nonet classification are possible:
\begin{description}
\item[1.] $f_0^{bare}(720)$, $f_0^{bare}(1260)$
\hfill 1$^3P_0$ nonet,\hfill
$f_0^{bare}(1600)$, $f_0^{bare}(1810)$
\hfill 2$^3P_0$ nonet;
\item[2.] $f_0^{bare}(720)$, $f_0^{bare}(1260)$
\hfill 1$^3P_0$ nonet,\hfill
$f_0^{bare}(1230)$, $f_0^{bare}(1810)$
\hfill 2$^3P_0$ nonet;
\end{description}
Scalar mesons in the lightest $1^3P_0~q\bar q$-nonet are the same in
both our solutions: $f_0^{bare}(720\pm 100)$ and $f_0^{bare}
(1260\pm 30)$. The flavour content of $f_0^{bare}(720\pm 100)$ and
$f_0^{bare}(1260\pm 30)$ almost coinsides with the $n\bar n/s\bar s$
content of $\eta$ and $\eta'$, correspondingly, that indicates
the symmetry in
interactions which are responsible for the formation of the lightest
scalar/pseudoscalar $q\bar q$-mesons.

In any variant one of the bare states, either $f_0^{bare}(1230)$ or
$f_0^{bare}(1600)$, is superfluous for the
$q\bar q$ systematics, and its coupling constants are in accordance with
the relations for glueball decay.

It should be emphasized that our bare state
does not correspond to a pure gluodynamic glueball of refs.
\cite{ukqcd,ibm}: the bare state includes
quark degrees of freedom, in particular
the $q\bar q$ component (this problem is discussed in detail in ref.
\cite{aas}). This means that the mass of
$f_0^{bare}(1230)$ or $f_0^{bare}(1600)$
should not coincide
with the mass obtained in Lattice calculations for the pure glueball:
$1550\pm 50$ MeV \cite{ukqcd} and $1710\pm 40$ MeV \cite{ibm}. Quite the
opposite, as is shown in ref. \cite{aas}, an admixture of
the $q\bar q$ component wants
to reduce the mass of a pure glueball by 200-300 MeV.  Therefore,
according to our fit, one would expect the mass of the gluodynamic
glueball in the region 1450-1600 MeV for the variant {\bf{1}}
or in the region 1700-1900 MeV for the variant {\bf{2}}.

A principal question to our analysis is how many  states are
in the region 1200-1400 MeV. We have investigated the variant with only one
bare state in this region: it makes the quality of the
fit worse.  The fit of the data set is based on the existence of three resonances
(amplitude poles) in the region 1200-1600 MeV: two comparatively narrow,
$f_0(1300)$ and $f_0(1500)$, and  a broad one, $f_0(1530^{+90}_{-250})$; the
interference of the broad resonance with narrow ones produces a wide variety
of effects which are typical for the spectra investigated.
In order to
have these three resonances we need three bare states in the region 1200-1600
MeV.

Nevertheless, one may be sceptical about taking into
consideration such a broad resonance like
$f_0(1530^{+90}_{-250})$. Here
we would like to emphasize the existence of a very important
phenomenon for overlapping resonances \cite{aas}: the mixing of
these resonances increases the width of one state, thus
simultaneously reducing the width of another one. In the case of
the full overlap, the width of one state tends to zero while
the width of the second state  tends to the sum of the widths of
initial states, $\Gamma_1+\Gamma_2$.
For three overlapping resonances, the width of two states tend to
zero while the width of the third state accumulates the widths of
all initial resonances, $\Gamma_{third}\simeq\Gamma_1+\Gamma_2+
\Gamma_3$.
This effect is quite similar
to what is well known in atomic/molecular physics, namely, the
repulsion of close levels. Here, in the case of poles in the
complex plane, the repulsion  has a tendency to
increase/decrease the widths. This means that in the case of
overlapping and mixing resonances it is inevitable to have at
least one resonance with a large width. Our analysis shows that
this effect happens exactly in the mass region 1200-1600 MeV.

\vskip 0.5cm
{\bf Conclusion}

We have performed a simultaneous analysis of all available data for
the $00^{++}$ channel in the mass region up to 1900 MeV.  Five bare states
are found: four of them are members of $q\bar q$-nonets, while one state
is extra for $q\bar q$ systematics: its couplings to meson
channels point out that it is a bare glueball state. This bare glueball
state is dispersed over three real resonances, $f_0(1300)$, $f_0(1500)$,
and $f_0(1530^{+90}_{-250})$.

\vskip 0.5cm
{\bf Acknowledgements}

We thank A.V.Anisovich, D.V.Bugg, L.G.Dakhno, S.S.Gershtein, A.A.Kondashov,
\\ A.K.Likhoded, L.Montanet and S.A.Sadovsky for useful discussions.
VVA and AVS are
grateful to the Russian Foundation for Fundamental Investigations (Grant N
96-02-17934) for  financial support. AVS is grateful to the fellowship
of INTAS grant 93-2492-ext; the work was carried out within the
research program of International Centre for Fundamental Physics in
Moscow.

\newpage
\begin{description}
\item [\bf Fig. 1.]
Quark-antiquark loop diagram which determines the glueball width
(a); diagrams for the decay of a
$q\bar q$-meson (b) and a glueball (c,d)
into two $q\bar q$-meson states.
\item [\bf Fig. 2.] The $\pi\pi\to \pi\pi$ $S$-wave amplitude
module squared
\cite{gams1}; the events are collected at the momentum transfer
squared $|t|<0.20$ GeV$^2$/c$^2$.
The solid curve corresponds to solution {\bf{II-2}}, the dashed curve
to  solution {\bf{I}} and the dotted one to  solution {\bf{II-1}}.
\item [\bf Fig. 3.] The $\pi\pi\to K\bar K$ $S$-wave amplitude squared:
data are taken from refs.\cite{bnl};
the style of the curves is the same as in Fig. 2.
\item [\bf Fig. 4.] The $\pi\pi\to \eta\eta$ $S$-wave amplitude
squared \cite{gams2},
the style of the curves is the same as in Fig. 2.
\item [\bf Fig. 5.] The $\pi\pi\to \eta\eta'$ $S$-wave amplitude
squared \cite{gams3},
the style of the curves is the same as in Fig. 2.
\item [\bf Fig. 6.] Event
numbers {\it versus}  invariant mass of the $\pi\pi$-system for different
$t$-intervals in the $\pi^-p\rightarrow \pi^0\pi^0n$ reaction
\cite{gams1}. The solid curves correspond to solution {\bf{II-2}}
and the dashed curves to solution {\bf{I}}.
\item [\bf Fig. 7.]
 Fit of the $\pi\pi$ angular-moment
distributions   in the final state of the reaction
$\pi^-p \to n\pi^+\pi^-$ at 17.2 GeV/c \cite{cern}. The curve
corresponds to solution {\bf{II-2}}.
\item [\bf Fig. 8.] $\pi^0\pi^0$ spectra: in the $p\bar p\to
\pi^0\pi^0\pi^0$ reaction (a), in the $p\bar p\to
\eta\pi^0\pi^0$ reaction (b); $\eta\eta$ spectra in the $p\bar p\to
\pi^0\eta\eta$ reaction (c), $\eta\pi^0$ spectra in the $p\bar p\to
\pi^0\pi^0\eta$ reaction (d).
Curves correspond to solution {\bf{II-2}}.
\end{description}
\newpage
\begin{center}
Table 1\\
Coupling constants given by quark combinatorics for a $q\bar q$-meson
decaying into two pseudoscalar mesons in the leading terms
of the $1/N$ expansion
and for glueball decay in the next-to-leading terms
of the $1/N$ expansion.
$\Phi$ is the mixing angle for $n\bar n$ and $s\bar s$ states, and
$\Theta$ is the mixing angle for $\eta -\eta'$ mesons:
$\eta=n\bar n \cos\Theta-s\bar s \sin\Theta$ and
$\eta'=n\bar n \sin\Theta+s\bar s \cos\Theta$.
In eq.(2) $g_1=\frac{\sqrt 3}{2} g\;\cos\Phi$,
$g_2=g (\sqrt 2\sin\Phi+\sqrt \lambda\cos\Phi)/2$.
\vskip 0.5cm
\begin{tabular}{|c|c|c|c|}
\hline
~      &     ~                    &  ~                     &~           \\
~      & The $q\bar q$-meson decay&Glueball decay couplings&Identity   \\
~      & couplings in the         &in the next-to-      &   factor in  \\
Channel& leading terms of $1/N$   &leading terms of $1/N$ &phase space \\
~      & expansion (Fig. 1e)      &expansion (Fig. 1f)     &~   \\
~      &     ~                    &  ~                     &~           \\
\hline
~ & ~ & ~ & ~ \\
$\pi^0\pi^0$ &
$g\;\cos\Phi/\sqrt{2}$& 0 & 1/2  \\
~ & ~ & ~ & ~ \\
$\pi^+\pi^-$ & $g\;\cos\Phi/\sqrt{2}$ & 0 &  1   \\
~ & ~ & ~ & ~ \\
$K^+K^-$ & $g (\sqrt 2\sin\Phi+\sqrt \lambda\cos\Phi)/\sqrt 8 $ & 0 & 1 \\
~ & ~ & ~ & ~ \\
$K^0K^0$ & $g (\sqrt 2\sin\Phi+\sqrt \lambda\cos\Phi)/\sqrt 8 $ & 0 & 1 \\
~ & ~ & ~ & ~ \\
$\eta\eta$ &
$g\left (\cos^2\Theta\;\cos\Phi/\sqrt 2+\right .$\hfill
&$2g_G(\cos\Theta-\sqrt{\frac{\lambda}{2}}\sin\Theta )^2$ &
1/2 \\
~ &\hfill$\left . \sqrt{\lambda}\;\sin\Phi\;\sin^2\Theta\right )$ &~ &\\
~ & ~ & ~ & ~ \\
$\eta\eta'$ &
$g\sin\Theta\;\cos\Theta\left(\cos\Phi/\sqrt 2-\right .$\hfill
&$2g_G(\cos\Theta-\sqrt{\frac{\lambda}{2}}\sin\Theta)$
 & 1\\
~& \hfill $\left .\sqrt{\lambda}\;\sin\Phi\right ) $ &
\hfill$(\sin\Theta+\sqrt{\frac{\lambda}{2}}\cos\Theta)$&\\
~ & ~ & ~ & ~ \\
$\eta'\eta'$ &
$g\left(\sin^2\Theta\;\cos\Phi/\sqrt 2+\right .$\hfill
&$2g_G (\sin\Theta+\sqrt{\frac{\lambda}{2}}\cos\Theta )^2$ &
1/2 \\
~&\hfill $\left .\sqrt{\lambda}\;\sin\Phi\;\cos^2\Theta\right)$ & ~ &\\
~ & ~ & ~ & ~ \\
\hline
\end{tabular}
\end{center}

\newpage
\begin{center}
Table 2\\
$\chi^2$ values for the K-matrix solutions.
\vskip 0.5cm
\begin{tabular}{|l|c|c|c|c|}
\hline
~ & ~ &  ~ & ~ & ~ \\
~ &  solution I &  solution II-1 & solution II-2 & Number of\\
~ & ~ &  ~ & ~ & points\\
\hline
Crystal Barrel &~  &~ & ~ & ~ \\
data \cite{cbc1,cbc2}& ~ &~ & ~ & ~ \\
$p\bar p\to \pi^0\pi^0\pi^0$   & 1.57  & 1.53  & 1.52 & 1338\\
$p\bar p\to \pi^0\eta\eta$     & 1.59  & 1.63  & 1.60 & 1798\\
$p\bar p\to \pi^0\pi^0\eta$    & 1.52  & 1.58  & 1.62 & 1738\\
\hline
CERN-M\"unich \cite{cern} & ~ &~ & ~ &~ \\
 data                          & 1.82 & 1.88 & 1.88 & 705 \\
$\pi^+\pi^-\to\pi^+\pi^-$  & ~   &~ & ~ &~ \\
\hline
New S-wave &~   & ~  & ~ &~ \\
GAMS data \cite{gams1}        & 1.18 &  1.39 & 1.42 &  68 \\
$\pi^+\pi^-\to\pi^0 \pi^0$ &~   &~ & ~ &~ \\
\hline
t-dependent & ~ &~ & ~ &~  \\
GAMS data \cite{gams1} & ~ &~ & ~ &~ \\
$ 0.00<t<0.20$            & 2.79  & 2.87  & 3.19  & 21  \\
$ 0.30<t<1.00$            & 2.98  & 3.04  & 2.84  & 38  \\
$ 0.35<t<1.00$            & 1.40  & 1.43  & 1.39  & 38  \\
$ 0.40<t<1.00$            & 2.20  & 2.16  & 2.38  & 38  \\
$ 0.45<t<1.00$            & 1.50  & 1.42  & 1.55  & 38  \\
$ 0.50<t<1.00$            & 1.92  & 1.82  & 1.97  & 38  \\
\hline
GAMS data \cite{gams2,gams3} & ~  &~ & ~ &~ \\
$\pi\pi\to \eta\eta$      & 0.70  & 0.88 & 0.99  & 16 \\
$\pi\pi\to \eta\eta'$     & 0.38  & 0.52 & 0.37  & 8  \\
\hline
Brookhaven  &~  &~ & ~ &~ \\
data \cite{bnl}              & 0.80 &  0.69 & 0.61 &  35 \\
$\pi\pi\to K\bar K$ &  ~   &  ~    & ~    & ~   \\
\hline
\end{tabular}
\end{center}
\newpage
\begin{center}
Table 3\\
Masses, coupling constants (in GeV) and mixing angles (in degrees)
for the $f_0^{bare}$-resonances for solution I.
The errors reflect the boundaries for a satisfactory
description of the data.
\vskip 0.5cm
\begin{tabular}{|l|ccccc|}
\hline
~ & \multicolumn{5}{c|}{Solution I}  \\
\hline
~ & $\alpha=1$ &$\alpha=2$ & $\alpha=3$ & $\alpha=4$ & $\alpha=5$ \\
\hline
~         &~ & ~ & ~ & ~ & ~ \\
M              & $0.651^{+.120}_{-.030}$ &$1.219^{+.045}_{-.030}$ &
$1.255^{+.015}_{-.045}$ & $1.617^{+.010}_{-.045}$ &
$1.813^{+.040}_{-.040}$ \\
~         &~ & ~ & ~ & ~ & ~ \\
$g^{(\alpha)}$ &$1.432^{+.100}_{-.100}$ &$0.612^{+.050}_{-.200}$
&$0.955^{+.080}_{-.080}$ &
$0.567^{+.050}_{-.050}$ &$0.567^{+.050}_{-.050}$\\
~         &~ & ~ & ~ & ~ & ~ \\
$g_G$     &0 &$-0.120^{+.050}_{-.080}$ &  0 & 0 & 0 \\
~         &~ & ~ & ~ & ~ & ~ \\
$g_{5}^{(\alpha)}$& 0 &$0.874^{+.100}_{-.150}$ & 0 &
$0.661^{+.100}_{-.150}$ & $0.557^{+.100}_{-.100}$  \\
~         &~ & ~ & ~ & ~ & ~ \\
$\Phi_\alpha $  & -68.5$^{+15}_{-4}$   & 25.0$^{+15}_{-15}$
&16.5$^{+8}_{-8}$& -6.0$^{+15}_{-17}$ & 89$^{+5}_{-15}$\\
~         &~ & ~ & ~ & ~ & ~ \\
\hline
~ & $a=\pi\pi$ &$a=K\bar K$ & $a=\eta\eta$ & $a=\eta\eta'$ & $a=4\pi$ \\
\hline
~         &~ & ~ & ~ & ~ & ~ \\
$f_{1a} $ &$0.505^{+.100}_{-.100}$ &$ 0.056^{+.100}_{-.100}$ &
$0.494^{+.100}_{-.100}$ &$0.438^{+.100}_{-.100}$ &
$-0.160^{+.100}_{-.100}$ \\
\hline
~ &  \multicolumn{4}{c}{$f_{ba}=0\qquad b=2,3,4,5$} & ~\\
~         &~ & ~ & ~ & ~ & ~ \\
~ & \multicolumn{2}{|c}{$g^{(1)}_3=-0.185^{+0.045}_{-0.045}$} &
\multicolumn{2}{c}{$g^{(1)}_4=-0.250^{+0.100}_{-0.100}$} &
$s_0=5^{+\infty}_{-2.5}$ \\
\hline
~ & \multicolumn{5}{c|}{Pole position}  \\
II sheet      &~ & ~ & ~ & ~ & ~ \\
under $\pi\pi$&$1.012^{+.008}_{-.008}$&~&~&~&~ \\
and $4\pi$    &$-i0.033^{+.008}_{-.004}$&~&~&~&~ \\
cuts          &~ & ~ & ~ & ~ & ~ \\
\hline
IV sheet      &~ & ~ & ~ & ~ & ~ \\
under $\pi\pi$,& ~ &$1.301^{+.010}_{-.020}$&$1.504^{+.004}_{-.008}$
              &$1.443^{+.150}_{-.120}$& ~ \\
$4\pi$, $K\bar K$,&~&$-i0.108^{+.025}_{-.015}$ &$-i0.064^{+.008}_{-.008}$ &
              $-i0.553^{+.080}_{-.120}$& ~ \\
$\eta\eta$ cuts&~ & ~ & ~ & ~ & ~ \\
\hline
V sheet       &~ & ~ & ~ & ~ & ~ \\
under $\pi\pi$,& ~  & ~  & ~  & ~ &$1.810^{+.020}_{-.020}$ \\
$4\pi,~K\bar K$,&  ~ &  ~ &  ~ & ~ &$-i0.112^{+.010}_{-.030}$   \\
\hline
\end{tabular}
\end{center}
\newpage
\begin{center}
Table 4\\
Masses, coupling constants (in GeV) and mixing angles (in degrees)
for the $f_0^{bare}$-resonances for the solution II-1.
\vskip 0.5cm
\begin{tabular}{|l|ccccc|}
\hline
~ & \multicolumn{5}{c|}{Solution II-1}  \\
\hline
~ & $\alpha=1$ &$\alpha=2$ & $\alpha=3$ & $\alpha=4$ & $\alpha=5$ \\
\hline
~         &~ & ~ & ~ & ~ & ~ \\
M              & $0.651^{+.120}_{-.030}$ &$1.220^{+.050}_{-.050}$
&$1.252^{+.020}_{-.030}$ &
$1.572^{+.040}_{-.030}$ &$1.820^{+.030}_{-.040}$ \\
~         &~ & ~ & ~ & ~ & ~ \\
$g^{(\alpha)}$ &$1.454^{+.100}_{-.150}$ &$0.605^{+.050}_{-.200}$
&$0.969^{+.080}_{-.080}$ &
$0.431^{+.050}_{-.050}$ &$0.431^{+.050}_{-.050}$\\
~         &~ & ~ & ~ & ~ & ~ \\
$g_G$     &0 & $-0.125^{+.050}_{-.080}$ &  0 & 0 & 0 \\
~         &~ & ~ & ~ & ~ & ~ \\
$g_{5}^{(\alpha)}$& 0 & $0.765^{+.100}_{-.150}$ & 0 &
$0.570^{+.100}_{-.100}$ & $-0.604^{+.120}_{-.120}$  \\
~         &~ & ~ & ~ & ~ & ~ \\
$\Phi_\alpha $  & -67.6$^{+15}_{-4}$   & 25.0$^{+25}_{-15}$
&17.4$^{+8}_{-8}$& 23.8$^{+15}_{-17}$ &-61.2$^{+15}_{-15}$\\
~         &~ & ~ & ~ & ~ & ~ \\
\hline
~ & $a=\pi\pi$ &$a=K\bar K$ & $a=\eta\eta$ & $a=\eta\eta'$ & $a=4\pi$ \\
\hline
~         &~ & ~ & ~ & ~ & ~ \\
$f_{1a} $ &$0.626^{+.100}_{-.200}$ &$-0.016^{+.100}_{-.100}$ &
$0.463^{+.100}_{-.200}$ &$0.496^{+.100}_{-.200}$ &
$-0.072^{+.150}_{-.100}$ \\
\hline
~ &  \multicolumn{4}{c}{$f_{ba}=0\qquad b=2,3,4,5$} & ~\\
~         &~ & ~ & ~ & ~ & ~ \\
~ & \multicolumn{2}{|c}{$g^{(1)}_3=-0.148^{+0.050}_{-0.050}$} &
\multicolumn{2}{c}{$g^{(1)}_4=-0.268^{+0.100}_{-0.100}$} &
$s_0=5^{+\infty}_{-2.5}$ \\
\hline
~ & \multicolumn{5}{c|}{Pole position}  \\
II sheet      &~ & ~ & ~ & ~ & ~ \\
under $\pi\pi$&$1.010^{+.008}_{-.008}$&~&~&~&~ \\
and $4\pi$    &$-i0.040^{+.006}_{-.008}$&~&~&~&~ \\
cuts          &~ & ~ & ~ & ~ & ~ \\
\hline
IV sheet      &~ & ~ & ~ & ~ & ~ \\
under $\pi\pi$,& ~ &$1.302^{+.010}_{-.020}$&$1.495^{+.006}_{-.006}$
              &$1.530^{+.100}_{-.200}$& ~ \\
$4\pi$, $K\bar K$,&~&$-i0.117^{+.015}_{-.025}$ &$-i0.061^{+.008}_{-.008}$ &
              $-i0.585^{+.050}_{-.100}$& ~ \\
$\eta\eta$ cuts&~ & ~ & ~ & ~ & ~ \\
\hline
V sheet       &~ & ~ & ~ & ~ & ~ \\
under $\pi\pi$,& ~  & ~  & ~  & ~ &$1.798^{+.020}_{-.020}$ \\
$4\pi,~K\bar K$,&  ~ &  ~ &  ~ & ~ &$-i0.089^{+.030}_{-.040}$   \\
$\eta\eta$ and &~ & ~ & ~ & ~ & ~ \\
$\eta\eta'$ cuts&~ & ~ & ~ & ~ & ~ \\
\hline
V sheet       &~ & ~ & ~ & ~ & ~ \\
under $\pi\pi$,& ~  & ~  & ~  & ~ &$1.798^{+.020}_{-.020}$ \\
$4\pi,~K\bar K$,&  ~ &  ~ &  ~ & ~ &$-i0.089^{+.030}_{-.040}$   \\
$\eta\eta$ and &~ & ~ & ~ & ~ & ~ \\
$\eta\eta'$ cuts&~ & ~ & ~ & ~ & ~ \\
\hline
\end{tabular}
\end{center}

\newpage
\begin{center}
Table 5\\
Masses, coupling constants (in GeV) and mixing angles (in degrees)
for the $f_0^{bare}$-resonances for the solution II-2.
\vskip 0.5cm
\begin{tabular}{|l|ccccc|}
\hline
~ & \multicolumn{5}{c|}{Solution II-2}  \\
\hline
~ & $\alpha=1$ &$\alpha=2$ & $\alpha=3$ & $\alpha=4$ & $\alpha=5$ \\
\hline
~         &~ & ~ & ~ & ~ & ~ \\
M              & $0.651^{+.120}_{-.030}$ &$1.219^{+.060}_{-.050}$
&$1.251^{+.020}_{-.030}$ &
$1.559^{+.060}_{-.020}$ &$1.821^{+.030}_{-.040}$ \\
~         &~ & ~ & ~ & ~ & ~ \\
$g^{(\alpha)}$ &$1.503^{+.100}_{-.200}$ &$0.508^{+.060}_{-.120}$
&$1.002^{+.060}_{-.100}$ &
$0.398^{+.070}_{-.040}$ &$0.508^{+.060}_{-.120}$\\
~         &~ & ~ & ~ & ~ & ~ \\
$g_G$     &0 & 0 & 0 & $0.030^{+.040}_{-.030}$&  0 \\
~         &~ & ~ & ~ & ~ & ~ \\
$g_{5}^{(\alpha)}$& 0 & $0.673^{+.120}_{-.100}$ & 0 &
$0.528^{+.100}_{-.100}$ & $-0.584^{+.120}_{-.120}$  \\
~         &~ & ~ & ~ & ~ & ~ \\
$\Phi_\alpha $  & -66.7$^{+15}_{-4}$   & 42.3$^{+8}_{-25}$
&18.3$^{+4}_{-8}$& 25.0$^{+5}_{-20}$ &-52.7$^{+10}_{-20}$\\
~         &~ & ~ & ~ & ~ & ~ \\
\hline
~ & $a=\pi\pi$ &$a=K\bar K$ & $a=\eta\eta$ & $a=\eta\eta'$ & $a=4\pi$ \\
\hline
~         &~ & ~ & ~ & ~ & ~ \\
$f_{1a} $ &$0.524^{+.150}_{-.150}$ &$-0.058^{+.100}_{-.100}$ &
$0.413^{+.100}_{-.120}$ &$0.406^{+.150}_{-.100}$ &
$-0.178^{+.150}_{-.100}$ \\
\hline
~ &  \multicolumn{4}{c}{$f_{ba}=0\qquad b=2,3,4,5$} & ~\\
~         &~ & ~ & ~ & ~ & ~ \\
~ & \multicolumn{2}{|c}{$g^{(1)}_3=-0.167^{+0.100}_{-0.100}$} &
\multicolumn{2}{c}{$g^{(1)}_4=-0.251^{+0.100}_{-0.100}$} &
$s_0=5^{+\infty}_{-2.5}$ \\
\hline
~ & \multicolumn{5}{c|}{Pole position}  \\
II sheet      &~ & ~ & ~ & ~ & ~ \\
under $\pi\pi$&$1.012^{+.008}_{-.008}$&~&~&~&~ \\
and $4\pi$    &$-i0.033^{+.008}_{-.004}$&~&~&~&~ \\
cuts          &~ & ~ & ~ & ~ & ~ \\
\hline
IV sheet      &~ & ~ & ~ & ~ & ~ \\
under $\pi\pi$,& ~ &$1.301^{+.010}_{-.020}$&$1.504^{+.004}_{-.008}$
              &$1.443^{+.150}_{-.120}$& ~ \\
$4\pi$, $K\bar K$,&~&$-i0.108^{+.025}_{-.0150}$ &$-i0.064^{+.008}_{-.008}$ &
              $-i0.553^{+.080}_{-.120}$& ~ \\
$\eta\eta$ cuts&~ & ~ & ~ & ~ & ~ \\
\hline
V sheet       &~ & ~ & ~ & ~ & ~ \\
under $\pi\pi$,& ~  & ~  & ~  & ~ &$1.814^{+.015}_{-.025}$ \\
$4\pi,~K\bar K$,&  ~ &  ~ &  ~ & ~ &$-i0.113^{+.010}_{-.030}$   \\
$\eta\eta$ and &~ & ~ & ~ & ~ & ~ \\
$\eta\eta'$ cuts&~ & ~ & ~ & ~ & ~ \\
\hline
\end{tabular}
\end{center}

\newpage
\begin{center}
Table 6 \\
The parameters of the $\pi\pi,~\eta\eta$ and $\eta\eta'$ production
amplitude $A_{\pi N\to N b}$ and $p\bar p$ annihilation amplitude
$A_k(s_{ij})$ for solution II-2.
All values are given in GeV.
\vskip 0.5cm
\begin{tabular}{|c|ccccc|}
\hline
~ & ~ & ~ & ~ & ~ & ~  \\
~ & \multicolumn{5}{c|}{$A_{\pi N\to N b}$}  \\
~ & ~ & ~ & ~ & ~ & ~  \\
\hline
~ & ~ & ~ & ~ & ~ & ~  \\
~ & $\alpha=1$ &$\alpha=2$ & $\alpha=3$ & $\alpha=4$ & $\alpha=5$ \\
\hline
~ & ~ & ~ & ~ & ~ & ~  \\
$g'^{(\alpha)}$ &-0.027 & 0     & 0.019 &  0.016 & 0 \\
\hline
~ & ~ & ~ & ~ & ~ & ~  \\
~ & $a=\pi\pi$ &$a=K\bar K$ & $a=\eta\eta$ & $a=\eta\eta'$ & $a=4\pi$ \\
\hline
~ & ~ & ~ & ~ & ~ & ~  \\
$f'_{a} $ &-0.025 & 0.027 & 0 & 0 & 0 \\
\hline
~ & ~ & ~ & ~ & ~ & ~  \\
& ~ & $N=474$ & $\tilde \Lambda=0.204$ &$\Lambda_g=2.46$&
~  \\
~ & ~ & ~ & ~ & ~ & ~  \\
\hline
\hline
~ & ~ & ~ & ~ & ~ & ~  \\
~ & \multicolumn{5}{c|}{$A_k(s_{ij})$}  \\
~ & ~ & ~ & ~ & ~ & ~  \\
\hline
~ & $\alpha=1$ &$\alpha=2$ & $\alpha=3$ & $\alpha=4$ & $\alpha=5$ \\
\hline
$Re(\Lambda^{(\alpha)}_{p\bar p\pi})$ & 0.023 & 0.590 & 0.389 & 1
&-0.100  \\
$Im(\Lambda^{(\alpha)}_{p\bar p\pi})$ &-0.387 &-0.016 &-0.430 & 0
&-0.192  \\
$Re(\Lambda^{(\alpha)}_{p\bar p\eta})$ & 1  &-0.304 &-0.171 & 0
& 0       \\
$Im(\Lambda^{(\alpha)}_{p\bar p\eta})$ & 0  & 0.243 & 0.473 & 0
& 0     \\
\hline
~ & ~ & ~ & ~ & ~ & ~  \\
~ & $a=\pi\pi$ &$a=K\bar K$ & $a=\eta\eta$ & $a=\eta\eta'$ & $a=4\pi$ \\
\hline
$Re(\phi_{p\bar p\pi,a})$ &-0.102 &-0.190 & 0.071 &0
& 0     \\
$Im(\phi_{p\bar p\pi,a})$ &-0.148 & 0.093 & 0.092 &0
& 0     \\
$Re(\phi_{p\bar p\eta,a})$ & 0.879 & 0.049 & 0     &0
& 0     \\
$Im(\phi_{p\bar p\eta,a})$ & 1.312 &-1.558 & 0     &0
& 0     \\
\hline
\end{tabular}
\end{center}

\newpage
\begin{center}
\epsfxsize=15cm \epsfbox{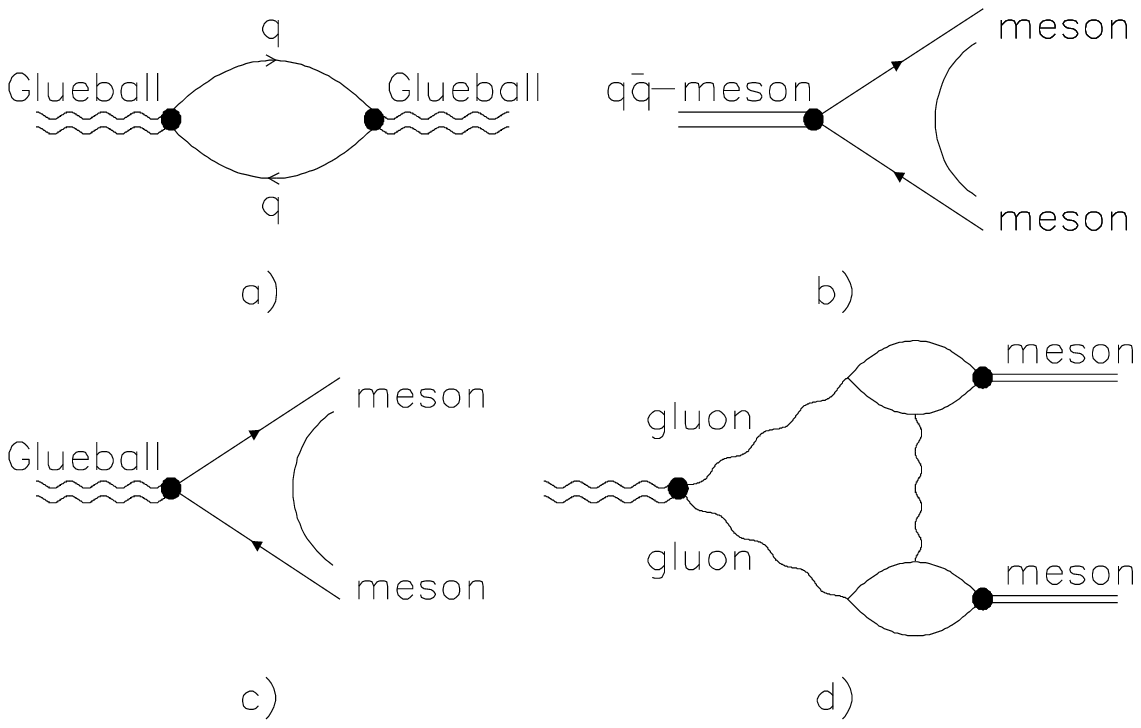}
Fig. 1
\end{center}

\begin{center}
\epsfxsize=15cm \epsfbox{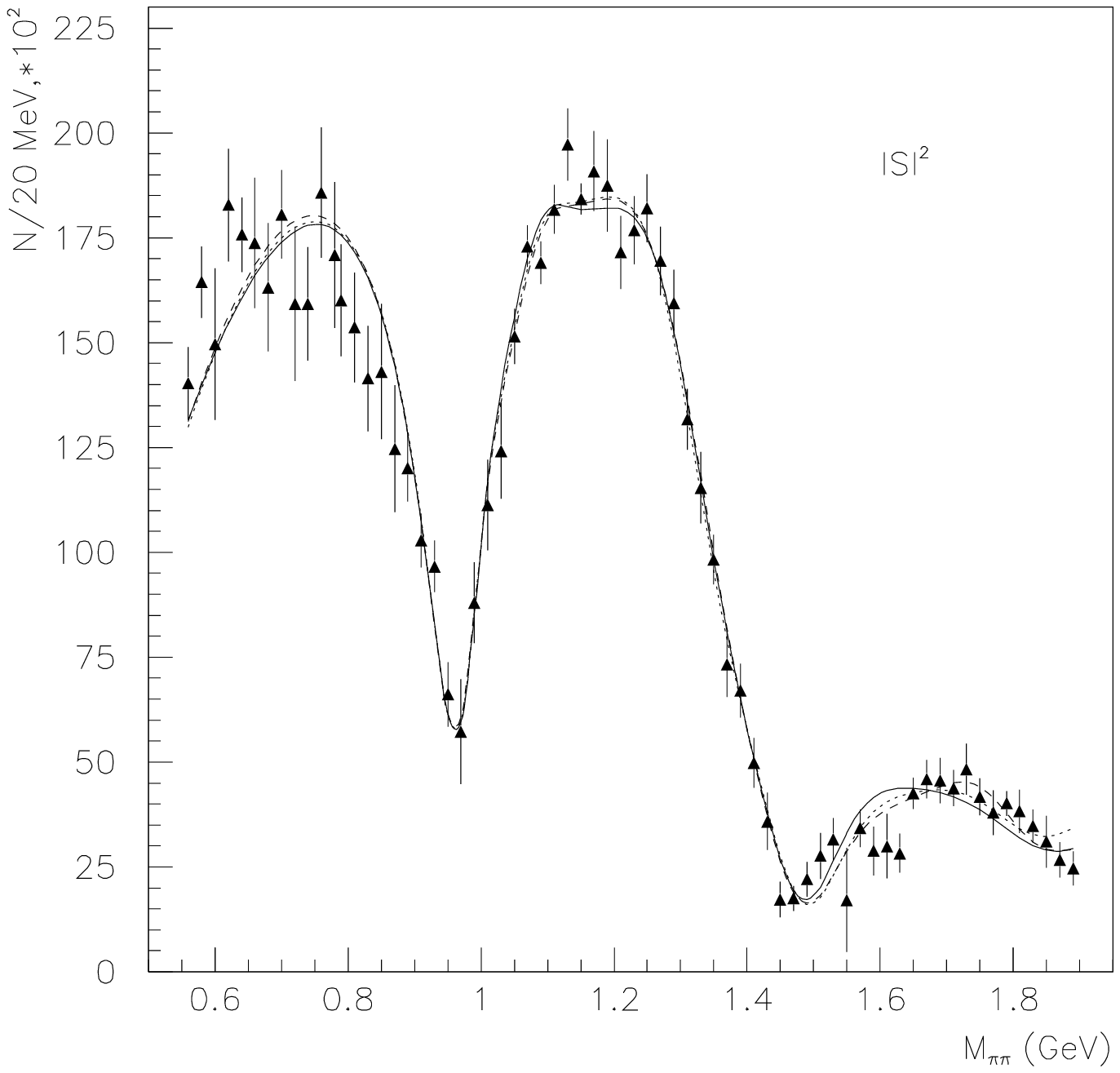}
Fig. 2
\end{center}

\begin{center}
\epsfxsize=15cm \epsfbox{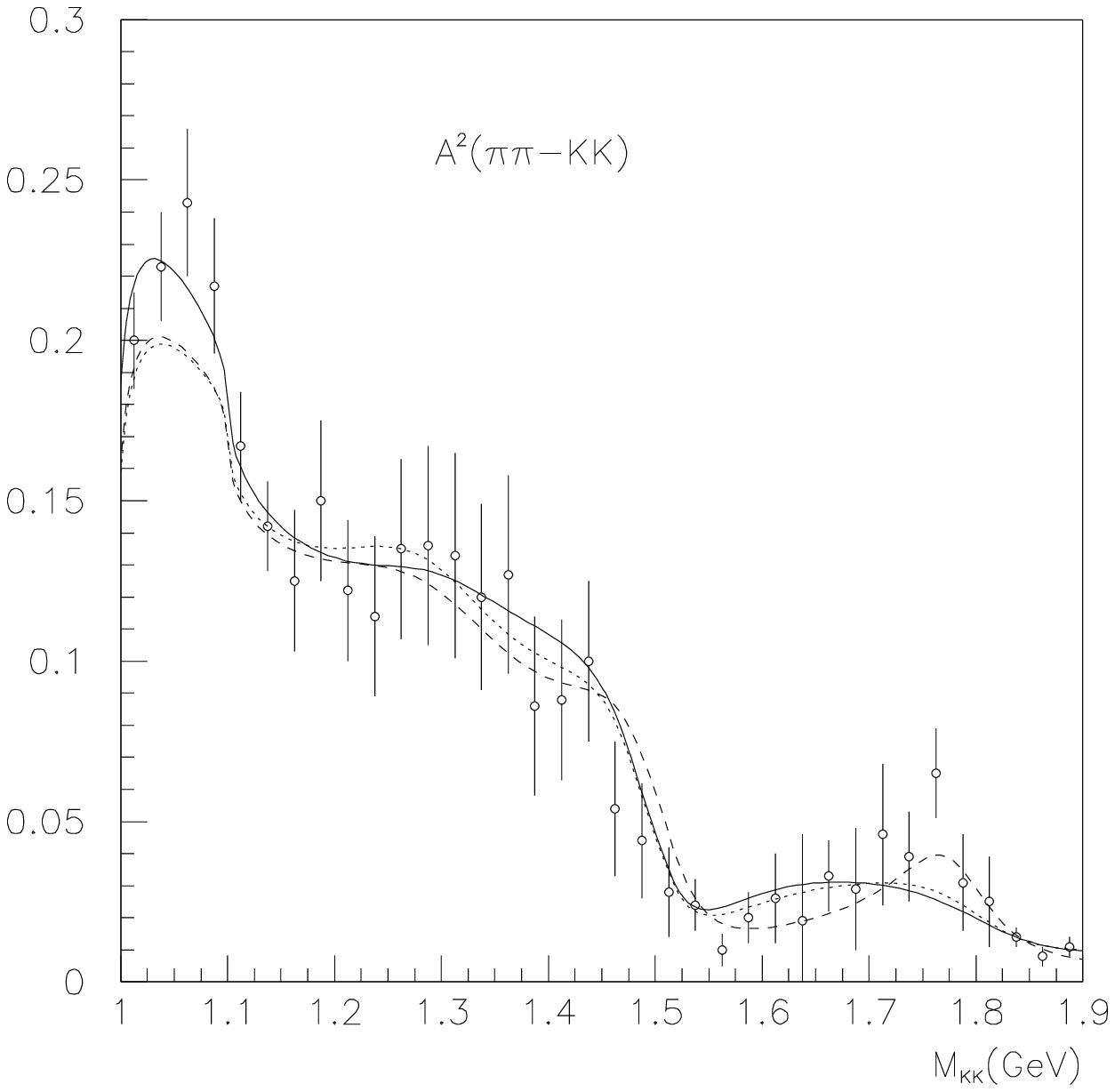}
Fig. 3
\end{center}

\begin{center}
\epsfxsize=15cm \epsfbox{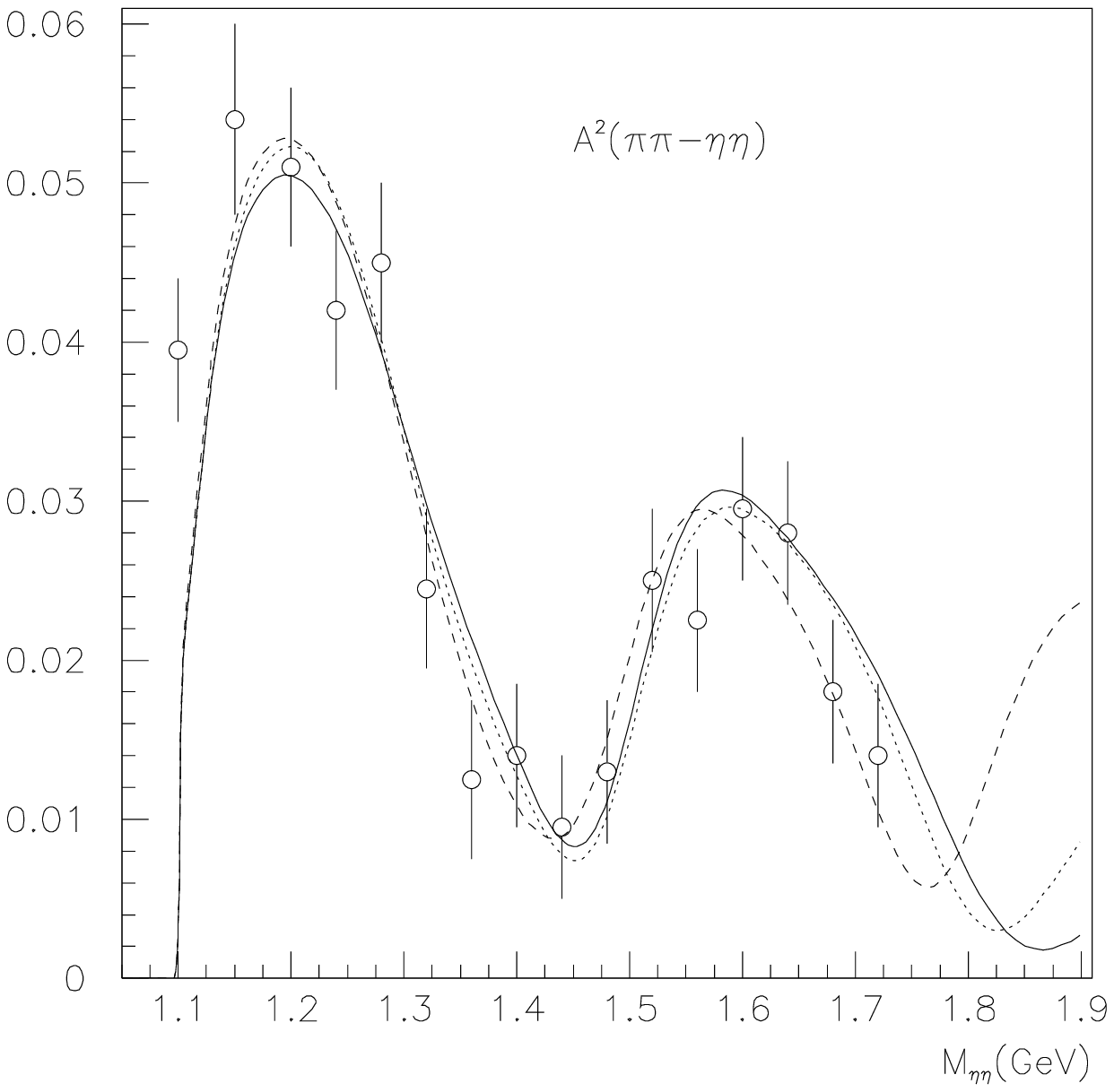}
Fig. 4
\end{center}

\begin{center}
\epsfxsize=12cm \epsfbox{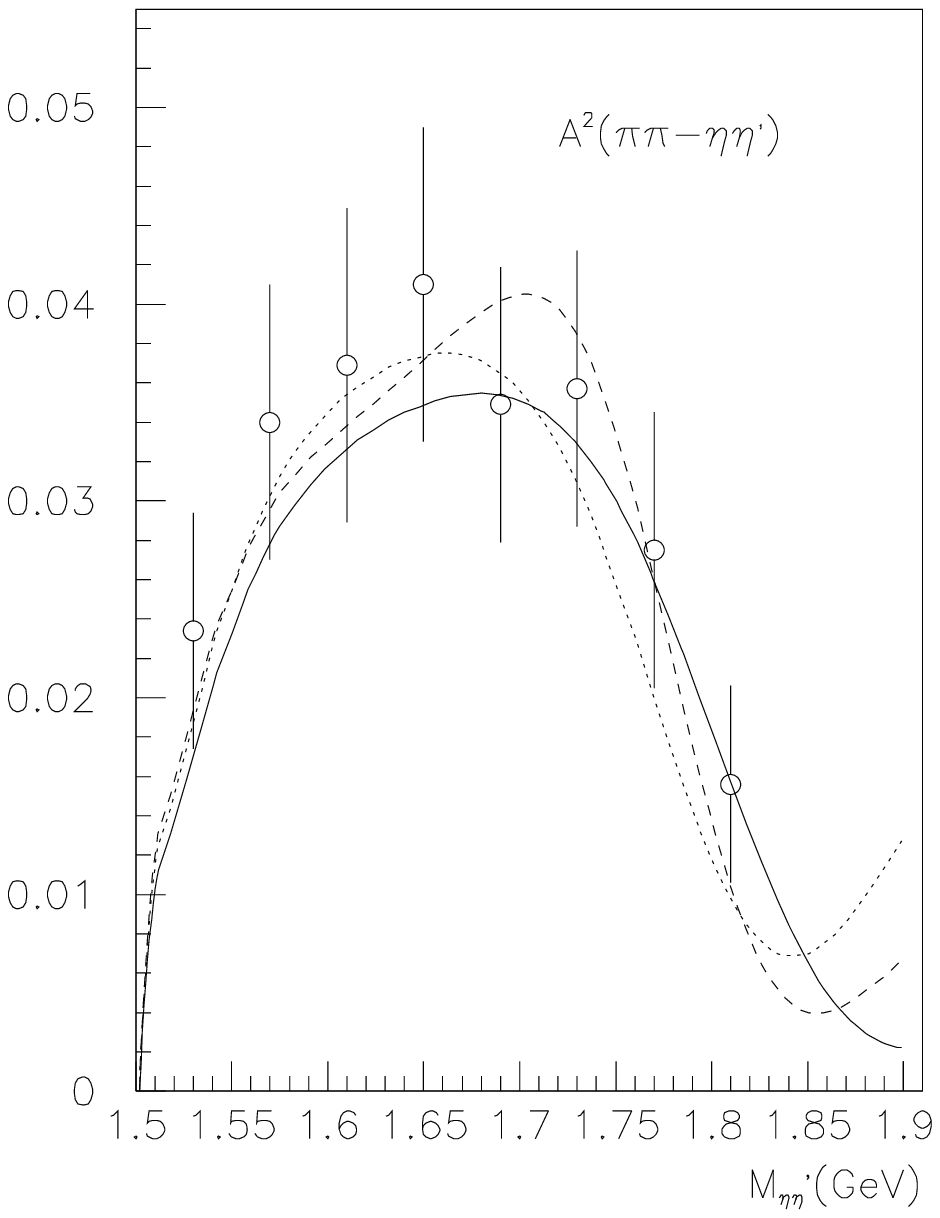}
Fig. 5
\end{center}

\begin{center}
\epsfxsize=15cm \epsfbox{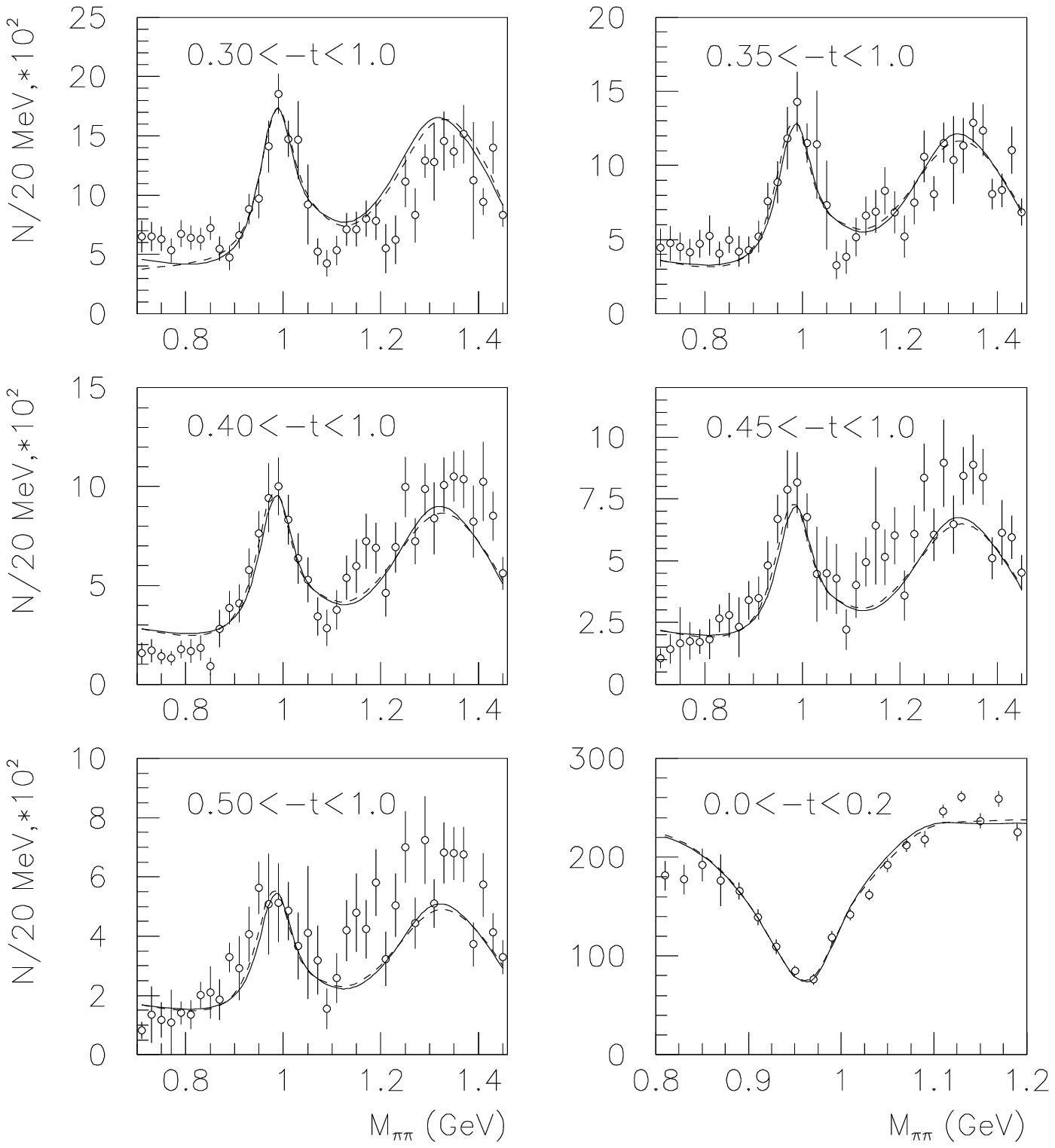}
Fig. 6
\end{center}

\begin{center}
\epsfxsize=15cm \epsfbox{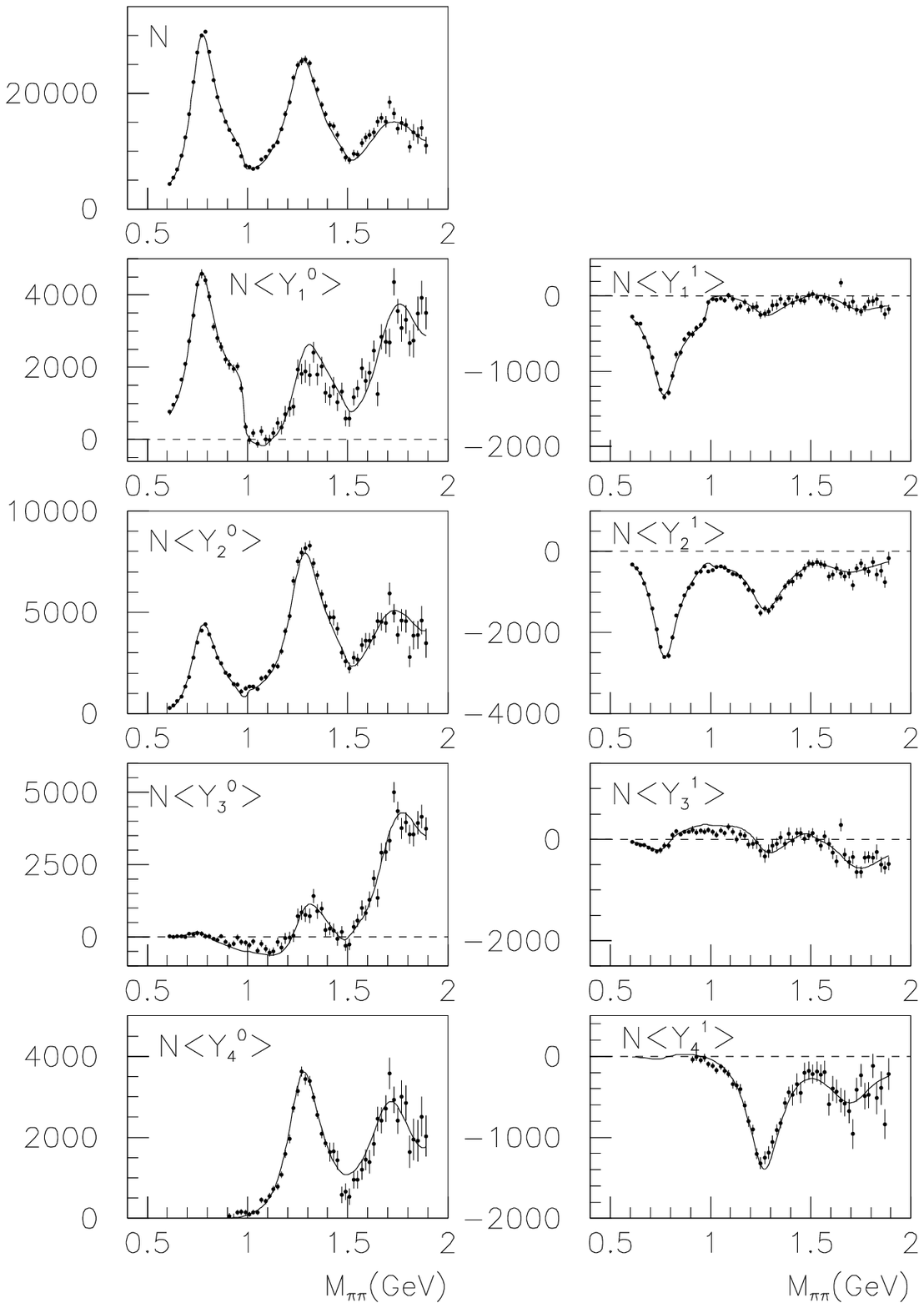}
Fig. 7
\end{center}

\begin{center}
\epsfxsize=15cm \epsfbox{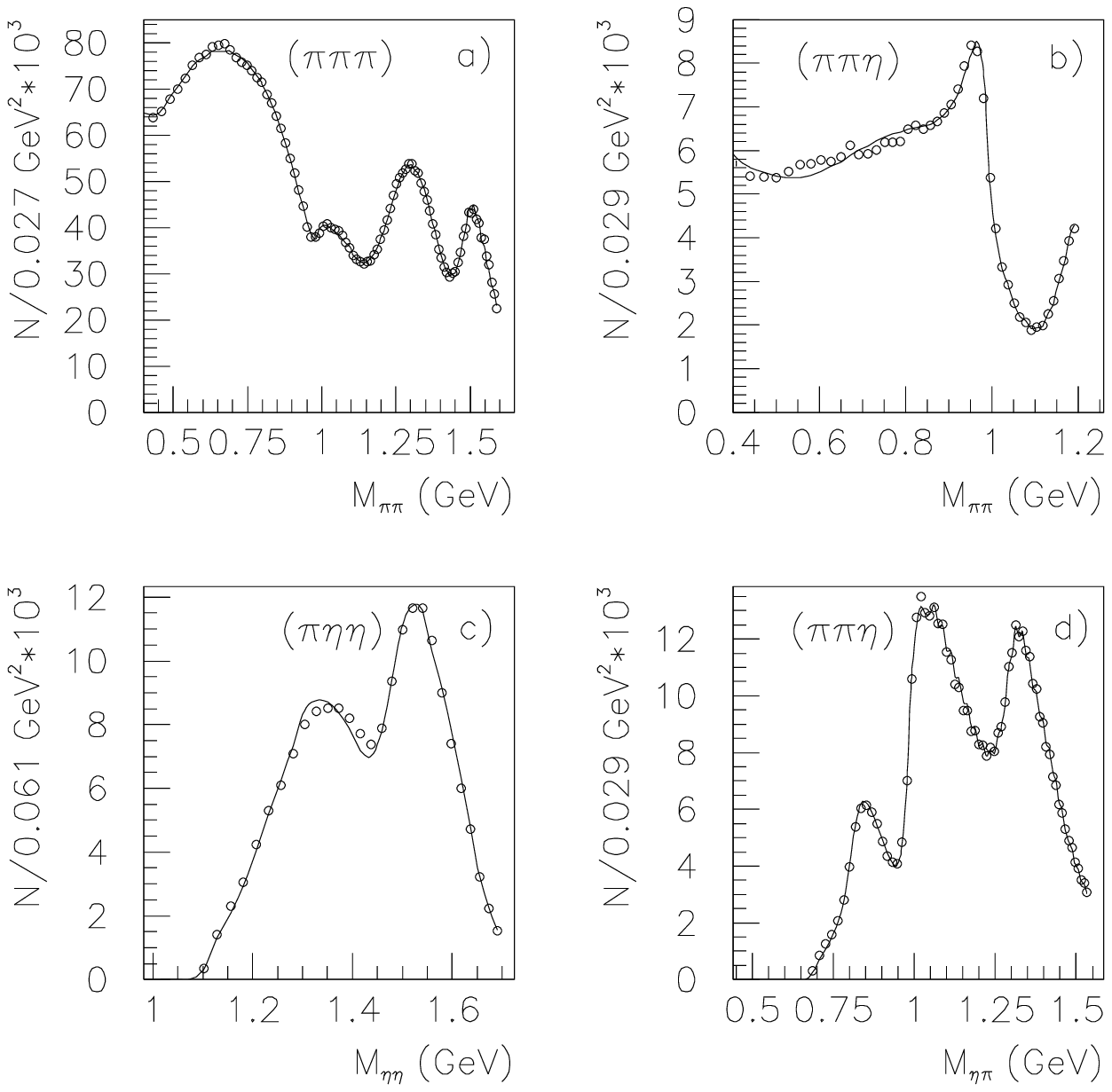}
Fig. 8
\end{center}

\newpage
\begin{thebibliography} {99}
\bibitem{gams1}
D. Alde et al., Z. Phys.{\bf C66} (1995) 375;\\
A.A. Kondashov et al., Proc. 27th Intern.
Conf. on High Energy Physics, Glasgow (1994) 1407;\\
Yu.D. Prokoshkin et al., Physics-Doklady {\bf 342} (1995), 473;\\
A.A. Kondashov et al, Preprint IHEP 95-137, Protvino (1995).
\bibitem{gams2} F. Binon et al., Nuovo Cim. {\bf A78} (1983) 313.
\bibitem{gams3} F. Binon et al., Nuovo Cim. {\bf A80} (1984) 363.
\bibitem{cbc1} V.V. Anisovich et al., Phys. Lett.
         {\bf B323} (1994) 233.
\bibitem{cbc2} C. Amsler et al., Phys. Lett. {\bf B342} (1995) 433.
\bibitem{cern} B. Hyams et al., Nucl. Phys. {\bf B64} (1973) 134.
\bibitem{bnl} S.J. Lindenbaum and R.S. Longacre, Phys. Lett.
{\bf B274} (1992) 492;\\
A. Etkin et al., Phys. Rev. {\bf D25} (1982) 1786.
\bibitem{aps} V.V. Anisovich et al., Phys. Lett. {\bf B355}
(1995) 363.
\bibitem{kmatr} V.V. Anisovich and A.V. Sarantsev,
Phys. Lett. B., in press.
\bibitem{bsz} D.V. Bugg, A.V. Sarantsev and B.S. Zou,
          Nucl. Phys. {\bf B}, in press.
\bibitem{pdg}  L. Montanet et al. (Particle Data Group),
Phys. Rev. {\bf D} 50 (1994) 1173.
\bibitem{ukqcd} G.S. Bali et al., Phys. Lett. {\bf B309} (1993)
378;\\ R. Gupta et al., Phys. Rev. {\bf D43} (1991) 2301.
\bibitem{ibm} J. Sexton, A. Vassarino and D. Weingarten,
Phys.Rev.Lett. {\bf 75} (1995) 4563.
\bibitem{aas} A.V. Anisovich, V.V. Anisovich and A.V. Sarantsev,
"The lightest glueball: Investigation of the $00^{++}$ wave in dispersion
relation approach", to be published.
\bibitem{gersh} S.S. Gershtein, A.K. Likhoded and
Yu.D. Prokoshkin, Z. Phys. {\bf C24} (1984) 305.
\bibitem{amsl} C. Amsler and F.E. Close, Phys. Lett. {\bf B353}
(1995) 385;
\bibitem{anis1} V.V. Anisovich, Phys. Lett. {\bf B364} (1995) 195.
\bibitem{thooft} G. t'Hooft, Nucl. Phys. {\bf B72} (1974) 461; \\
G. Veneziano, Nucl. Phys. {\bf B117} (1976) 519.
\bibitem{ahkm} V.V. Anisovich, M.G. Huber, M.N. Kobrinsky and B.Ch. Metch,
Phys. Rev. {\bf D42} (1990) 3045.
\bibitem{ansh} V.V. Anisovich and V.M. Shekhter, Nucl. Phys. {\bf B55}
(1973) 455.
\bibitem{bjfa} J.D. Bjorken and G.E. Farrar, Phys. Rev. {\bf D9}
(1974) 1449.
\bibitem{vol} M.A. Voloshin, Yu.P. Nikitin and P.I. Porfirov,
Yad. Fiz. {\bf 35} (1982) 1006;
[Sov. J. Nucl. Phys. 35 (1982) 586].
\bibitem{jpsi1} D.V. Bugg et al., Phys. Lett. {\bf 353} (1995) 378.
\bibitem{jpsi2} V.V. Anisovich et al., "Resonances in $J/\psi
\to \gamma(\pi^+\pi^-\pi^+\pi^-)$", Preprint PNPI-TH-59-1994-2001
(1994).
\end{thebibliography}
\end{document}